\documentclass[twocolumn,secnumarabic,amssymb,nobibnotes,aps,prl,superscriptaddress]{revtex4-2}

\usepackage[utf8]{inputenc}
\usepackage[dvipsnames, table]{xcolor}
\usepackage{subfigure}
\usepackage[no-test-for-array]{nicematrix}

\usepackage{amsmath, mathtools, acronym, multirow, booktabs}
\usepackage[per-mode = repeated-symbol]{siunitx}
\usepackage{tabularx}
\usepackage{adjustbox}
\usepackage{hyperref}
\usepackage{float}
\usepackage[normalem]{ulem}

\usepackage{graphicx}
\graphicspath{{./figures/}}
 
\hypersetup{
    colorlinks=true,
    linkcolor=blue,
    filecolor=magenta,      
    urlcolor=blue,
}

\DeclareMathAlphabet{\mathpzc}{OT1}{pzc}{m}{it}

\newcommand{\IFAE}{Institut de Física d’Altes Energies (IFAE), The Barcelona Institute of Science and Technology, UAB Campus, E-08193 Barcelona, Spain}
\newcommand{\ICREA}{Catalan Institution for Research and Advanced Studies (ICREA), E-08010 Barcelona, Spain}
\newcommand{\NIKHEF}{Nikhef, Science Park 105, 1098 XG Amsterdam, The Netherlands}
\newcommand{\UU}{Institute for Gravitational and Subatomic Physics (GRASP), \mbox{Utrecht University}, Princetonplein 1, 3584 CC Utrecht, The Netherlands}
\newcommand{\PISA}{Dipartimento di Fisica “Enrico Fermi”,  \mbox{Università di Pisa} and INFN Sezione di Pisa, Pisa I-56127, Italy}

\begin{document}

\title{First constraints on compact binary environments from LIGO-Virgo data}

\author{Giada Caneva Santoro}
\email{gcaneva@ifae.es}
\affiliation{\IFAE}

\author{Soumen Roy}
\email{soumen.roy@nikhef.nl}
\affiliation{\NIKHEF}
\affiliation{\UU}

\author{Rodrigo Vicente}
\email{rvicente@ifae.es}
\affiliation{\IFAE}

\author{Maria Haney}
\affiliation{\NIKHEF}

\author{Ornella Juliana Piccinni}
\affiliation{\IFAE}

\author{Walter Del Pozzo}
\affiliation{\PISA}

\author{Mario Martinez}
\affiliation{\IFAE}
\affiliation{\ICREA}

\definecolor{shadowcolor}{gray}{0.9}
\definecolor{noshadowcolor}{gray}{1}
\setlength{\tabcolsep}{0pt}
\renewcommand{\arraystretch}{1.3}

\def \msun	{{\rm M}_\odot}
\def \angle {55}
\def \imrpd {\texttt{IMRPhenomD}}
\def \imrpv {\texttt{IMRPhenomPv2}}
\def \imrpnrtv {\texttt{IMRPhenomPv2\_NRTidalv2}}
\def \lalsim {\textsc{LALSimulation}}
\def \lalinf {\textsc{LALInference}}
\def \bilby {\textsc{Bilby}}
\def \numpy {\textsc{NumPy}}
\def \scipy {\textsc{SciPy}}
\def \aligo {\textsc{aLIGO}}
\def \flow	{f_{\mathrm{low}}}
\def \fhigh	{f_{\mathrm{high}}}
\def \Hz	{\rm{Hz}}
\def \mismatch {\mathpzc{MM}}
\def \ff {\mathpzc{FF}}
\def \hvac {h_{\rm vac}}
\def \henv {h_{\rm env}}
\def \dfk {\delta\Phi_k}
\def \dfkhat {\widehat{\delta \Phi}_k}
\def \lnB {\log_{10}\mathcal{B}_{\rm vac}^{\rm env}}
\def \densityUnit {\si{\gram\per\cubic\centi\metre}}

\def \Henv {{\mathcal{H}}_{\rm env}}
\def \Hvac {{\mathcal{H}}_{\rm vac}}

\newcommand\inp[2]{\langle #1 \,|\, #2 \rangle}

\newcolumntype{C}{>{\centering\arraybackslash}X}

\date{\today}

\begin{abstract}
The LIGO-Virgo analyses of signals from compact binary mergers observed so far have assumed isolated binary systems in a vacuum, neglecting the potential presence of astrophysical environments. We present here the first investigation of environmental effects on each of the events of GWTC-1 and two low-mass events from GWTC-2.  We find no evidence for the presence of environmental effects. Most of the events decisively exclude the scenario of dynamical fragmentation of massive stars as their formation channel. GW170817 results in the most stringent upper bound on the medium density ($\lesssim 21\: \densityUnit$). We find that environmental effects can substantially bias the recovered parameters in the vacuum model, even when these effects are not detectable. We forecast that the Einstein Telescope and B-DECIGO will be able to probe the environmental effects of accretion disks and superradiant boson clouds on compact binaries.

\end{abstract}

\maketitle

{\bf \em Introduction.} 
%
In-depth analyses of gravitational-wave (GW) data have been routinely conducted by the LIGO-Virgo-KAGRA (LVK) collaboration, through parameter estimation, population studies, cosmology, and tests of general relativity (GR)~\cite{LIGOScientific:2018mvr, LIGOScientific:2020ibl, LIGOScientific:2021djp, LIGOScientific:2018jsj, LIGOScientific:2020kqk, KAGRA:2021duu, LIGOScientific:2017adf, LIGOScientific:2019zcs, LIGOScientific:2021aug, LIGOScientific:2021usb, LIGOScientific:2019fpa, LIGOScientific:2020tif, LIGOScientific:2021sio, KAGRA:2013rdx}. 
These analyses relied on the assumption that the sources of GWs are in a \emph{vacuum} environment.
However, there is a growing interest in exploring the potential effects of astrophysical environments on these observations, particularly with regard to binary black hole (BBH) formation in dense regions, such as star clusters~\cite{Banerjee:2010,Chatterjee:2017,Ziosi:2014sra,Rodriguez:2015oxa,Mapelli:2021gyv,Mapelli:2021taw}, active galactic nuclei (AGN) accretion disks~\cite{Sedda:2023big, Rowan:2022ehz, Ishibashi_2020}, or through the dynamical fragmentation of very massive stars~\cite{Loeb:2016fzn}. The physical processes that take place in environments that harbour black holes (BHs) significantly impact their formation, dynamics and evolution, and it is therefore crucial to characterise them. These effects are also relevant to multi-messenger astronomy, as electromagnetic counterparts are expected to accompany GWs for BBH mergers in non-vacuum environments~\cite{Loeb:2016fzn,McKernan:2019hqs,Perna:2019pzr,Graham:2020gwr}.

Most studies of environmental effects on GW signals (e.g.,~\cite{Kocsis:2011dr,Yunes:2011ws,Barausse:2014tra,Barausse:2014pra,Tamanini:2019usx,Toubiana:2020drf,Cardoso:2019rou,Derdzinski:2020wlw,Zwick:2021dlg,Kuntz:2021hhm,Sberna:2022qbn,Speri:2022upm,Cardoso:2022whc,Cole:2022yzw,Zwick:2022dih,Speeney:2022ryg,Kuntz:2022onu,Kuntz:2022juv,Boudon:2023vzl,Vijaykumar:2023tjg,Duque:2023cac}) have focused on sources relevant to the frequency bandwidth of the LISA experiment~\cite{Amaro-Seoane2017}, like intermediate and extreme mass ratio inspirals. These works forecast that LISA will be sensitive to the imprints of astrophysical environments, potentially being able to identify the type of environment and reconstruct its model parameters~\cite{Cole:2022yzw}. The reason for the generalized focus on LISA sources is mainly twofold:~(i) the long observation window of the inspiral stage, which allows for the accumulation of usually small environmental effects, and~(ii) LISA's sensitivity to low frequencies, which are the most affected by environmental dephasing. LISA is expected to become operational only by 2037~\cite{LISA_summary} and so it is important to quantify the \emph{current} ability of LVK to ``see'' environments.

Here, for the first time, a systematic Bayesian analysis is performed in the framework of the existing LIGO-Virgo data, testing for the presence of environmental effects, trying to constrain the environmental properties, and determining whether the inclusion of environmental effects in waveforms can bias the estimations of the GW source parameters. A previous study using LIGO-Virgo data focused on the first observed event GW150914 by comparing numerical relativity waveforms in vacuum versus environment~\cite{Fedrow:2017dpk}, which is computationally challenging and out of scope for low-mass binaries. 
We perform, for the first time, a parameterized post-Newtonian (PN) test of environmental effects with LIGO-Virgo data. The test allows for modifications to the in-vacuum GW phase of the binary that could be induced by environments during the inspiral stage.
Similar types of parameterized tests, considering waveform deformations at PN orders higher than the ones induced by environmental effects, are routinely performed by LVK to test GR~\cite{LIGOScientific:2019fpa,LIGOScientific:2020tif,LIGOScientific:2021sio}. The same framework can be used to detect non-GR signals~\cite{Narola:2022aob, Sharma:2023djw}. However, these methods do not specifically test scenarios of a binary immersed in an astrophysical medium.

We use units where $G=c=1$.
Compact binaries with stellar mass components may form and evolve in the vicinity of a third massive BH (MBH)~\cite{Tagawa:2019osr}. For reference, a MBH with mass $M_{\mathrm{MBH}}$ can lead to environments with densities as large as $(10^5 M_\odot/M_\mathrm{MBH})10^{-8}\,\mathrm{g}/\mathrm{cm}^{3}$ for thick accretion disks and $(10^5 M_\odot/M_\mathrm{MBH})^\frac{7}{10}\,0.1\,\mathrm{g}/\mathrm{cm}^{3}$ for thin ones~\cite{Barausse:2014tra},~$10^{-6}\,\mathrm{g}/\mathrm{cm}^3$ for dark matter spikes~\cite{Gondolo:1999ef,Sadeghian:2013laa}, and $(10^5 M_\odot/M_{\mathrm{MBH}})^2\,0.1\,\mathrm{g}/\mathrm{cm}^3$ for superradiant clouds of ultra-light bosons~\cite{Brito:2015oca}.
Compact binaries formed through the dynamical fragmentation of a massive star~\cite{Loeb:2016fzn} are expected to subsequently merge in a gaseous environment with density~$\gtrsim 10^7\,\mathrm{g}/\mathrm{cm}^3$~\cite{Fedrow:2017dpk}.

{\bf \em Environmental effects.} 
%
\label{sec:env-effects}
When in astrophysical environments, the phase evolution of compact binaries is expected to be slightly modified with respect to the vacuum ``chirp'' by effects like accretion~\cite{Bondi:1944jm,Bondi:1952ni,Petrich:1989} or dynamical friction (DF)~\cite{Chandrasekhar:1943ys,Ruderman:1971,Rephaeli:1980,Tremaine:1984,Ostriker:1998fa} (for details, see Ref.~\cite{Barausse:2007ph}). While the specifics of these effects and their relative impact on waveforms depend on the particular environment and binary source and can only be completely captured by numerical simulations, there are generic (agnostic) features that can be well captured by semi-analytic expressions. 

While environmental effects may lead to an increase in the BBH eccentricity~\cite{Cardoso:2020iji, Ishibashi_2020}, for small enough densities the GW radiation reaction dominates leading to the circularization of these binaries~\cite{Peters:1964zz}. For quasi-circular orbits, DF from the gravitational wake in the medium results in an energy loss
\begin{equation}\label{eq:Eloss_DF}
    \dot{E}_{\mathrm{DF}} \approx  \frac{4 \pi \rho M^2 \mathcal{I}(v, \eta)}{v}\Big(\frac{1-3 \eta}{\eta}\Big),
\end{equation}
with~$v\coloneqq (\pi M_z f)^{\frac{1}{3}}$, where~$f/2$ denotes the orbital frequency and we define~$M_z\coloneqq(1+z)M$, with~$M$ the binary's total mass (in the source frame) and~$z$ the cosmological redshift to the source. The symmetric mass ratio is defined as~$\eta\coloneqq m_1 m_2/M^2$,~$\rho$ is the (local) average mass density of the environment, and~$\mathcal{I}$ is a function of~$v$ and~$\eta$ which depends on the type of environment; sources in the LVK mass range in a gaseous media with asymptotic speed of sound~$c_\mathrm{s}$ have~$\mathcal{I}\sim O\big(c_\mathrm{s}/10^{-3}\big)$, and so we treat~$\mathcal{I}$ as a constant in our analysis. Analytic expressions for~$\mathcal{I}$ can be found, e.g., in~\cite{Desjacques:2021} for gaseous media, or~\cite{Baumann:2021fkf,Buehler:2022tmr,Vicente:2022ivh,Traykova:2023qyv,Tomaselli:2023ysb} for (ultra-light) dark matter. In this work we assume a static homogeneous density profile, neglecting the feedback of the BBH on the environment (see, e.g.,~\cite{Kavanagh:2020cfn}).

Accretion causes the masses of the binary components to change through the inspiral, which also affects the evolution of the orbital phase~\footnote{The transfer of momentum via accretion is also responsible for an additional (hydrodynamic) drag force which may be included within the factor~$\mathcal{I}$ in Eq.~\eqref{eq:Eloss_DF}~\cite{Petrich:1989,Barausse:2007dy,Traykova:2023qyv}.}. In collisional media, numerical simulations (e.g.,~\cite{Farris:2009mt,Khan:2018ejm,Comerford:2019,Antoni:2019pgq,Kaaz:2021xla}) show that binary accretion can be well approximated by Bondi-Hoyle-Lyttleton accretion (BHLA), with the usual fudge parameter~$\lambda \sim O(1)$~\cite{Edgar:2004mk}. In media constituted by particles with larger mean free path (like particle dark matter overdensities~\cite{Barausse:2014tra,Eda:2013gg}, or plasmas around BHs~\cite{Galishnikova:2022mjg}), accretion is less efficient and it is better described by collisionless accretion (CA)~\cite{Eddington:1988,Begelman:1977}.

The environment can alter the evolution of the GW phase also through other effects (like the medium's gravitational potential~\cite{Barausse:2006vt,Eda:2013gg,Macedo:2013qea}, acceleration of the binary's center of mass~\cite{Toubiana:2020drf,Cardoso:2020lxx}, planet-like migration~\cite{Kocsis:2011dr,Yunes:2011ws,Speri:2022upm}, and others), but for LVK binaries the effects imparted by DF and accretion are expected to be the most important~\cite{Barausse:2014tra,Cardoso:2019rou}. We consider diluted environments causing only small corrections to the dominant ``chirp'' due to vacuum radiation-reaction. As shown in \emph{Supplemental Material} (see also Ref.~\cite{Barausse:2014tra}), at \emph{linear order} in~$\rho M^2$, the GW phase in environments differs from the vacuum one by additive terms~$(3/128 \eta) \dfkhat v^{k-5}$, where~$k=-9$ ($-4.5$PN) for CA and $k=-11$ ($-5.5$PN) for DF and BHLA, with the coefficients given in terms of physical parameters in Tab.~\ref{tab:coeff}. The same~$-5.5$PN effect on the GW phase has also been observed in numerical relativity simulations of BBHs in gaseous media (see Fig.~2 in Ref.~\cite{Fedrow:2017dpk}). We have also checked that, up to a $O(1)$ factor, the coefficients~$\beta_k$ of Tab.~\ref{tab:coeff} can reproduce the environment effect on the chirp observed in those simulations. We neglect the effect of BH spin on the environmental effects, since these are mainly due to density-perturbations in the weak-field region~\cite{Shapiro:1974} (however, see also~\cite{Aguayo-Ortiz:2021jzv}).

\begin{table}
    \caption{Dependence of environmental dephasing coefficients on the physical parameters (see \emph{Supplemental Material}). We write~$\dfkhat=-\beta_k \tilde{\rho} M^2$, with~$\tilde{\rho}$ denoting~$\rho$ for CA,~$\lambda \rho$ for BHLA, and~$\mathcal{I \rho}$ for DF. The effects enter at~$k/2$-th PN order in the GW phase.}
    \label{tab:coeff}
    \begin{center}
    \noindent
    \begin{tabularx}{24em}{ @{} C C C @{} }
            \toprule
            Effect & $\beta_k$ & $k$ \\
            \midrule
            CA & $  \dfrac{125\pi(1-3\eta)}{357\eta^2} $ & $-9$ \\
            BHLA & $  \dfrac{125\pi[1-5\eta(1-\eta)]}{1824\eta^4} $ & $-11$ \\
            DF & $ \dfrac{25\pi(1-3\eta)}{304\eta^3} $ & $-11$ \\
            \bottomrule
    \end{tabularx}
    \end{center}
\end{table}

{\bf \em Analysis Setup.} 
%
We consider an agnostic dephasing parameter~$\dfk$, such that, in frequency domain, the GW phase with a specific environmental effect is 
\begin{equation}
    \phi^{\rm env} = \phi^{\rm{vac}}+\frac{3}{128\eta} \dfk v^{k-5},
\end{equation}
where~$k=-9$ for CA, and~$k=-11$ for BHLA or DF.
To generate the waveforms with the environmental correction, we followed the model-agnostic framework of parameterized tests of GR~\cite{Li:2011cg, Agathos:2013upa, Meidam:2017dgf}. 
Throughout this work, we use the phenomenological inspiral-merger-ringdown models~$\imrpv$ for binary black hole (BBH)~\cite{Husa:2015iqa,Khan:2015jqa, Hannam:2013oca, PhenomPv2} and~$\imrpnrtv$ for binary neutron star (BNS) systems~\cite{Dietrich:2019kaq}, which incorporate spin-induced precession effects.

\begin{table*}
  \centering
  \begin{tabular}{lccccccccccccc}
    \toprule
          &  \rotatebox[origin=c]{\angle}{GW150914} &   \rotatebox[origin=c]{\angle}{GW151012} & \rotatebox[origin=c]{\angle}{ GW151226} &   \rotatebox[origin=c]{\angle}{GW170104} &   \rotatebox[origin=c]{\angle}{GW170608} &   \rotatebox[origin=c]{\angle}{GW170729} &   \rotatebox[origin=c]{\angle}{GW170809} &  \rotatebox[origin=c]{\angle}{GW170814} &   \rotatebox[origin=c]{\angle}{GW170817} & \rotatebox[origin=c]{\angle}{GW170818} & \rotatebox[origin=c]{\angle}{GW170823} & \rotatebox[origin=c]{\angle}{GW190425} & \rotatebox[origin=c]{\angle}{GW190924} \\
    \midrule
    $\delta\Phi_{-9}$  &      $-2.09$ &      -- &      $-4.20$ &      $-1.29$ &      $-4.95$ &      -- &      $-1.91$ &      $-3.17$ &   $-5.45$ &   -- &      --  &  $-10$ & $-5.5$ \\
    $\delta\Phi_{-11}$ &      $-3.30$ &       -- &     $-5.52$ &      $-3.33$ &      $-6.17$ &      -- &      $-2.59$ &      $-2.81$ &    $-6.47$ &  -- &      --  & $-11.6$ & $-6.4$ \\
    \bottomrule
  \end{tabular}
  \caption{Logarithmic Bayes factor ($\lnB$) for LIGO-Virgo events. For the events GW170729 and GW170823, we could not find informative~$\dfk$ posteriors even with a broad prior range due to the low SNR of their inspiral. For GW151012 and GW170818, the sampler failed to converge in the sense that individual runs result in different posterior distributions.
  }
  \label{tab:bayesF}
\end{table*}

We carry out a Bayesian parameter estimation analysis to measure the dephasing parameters~$\dfk$ and quantify the evidence for the presence of an environment. We compute the Bayes factor~$\mathcal{B}_{\rm vac}^{\rm env}$ to compare the two hypotheses: (i) the data~$d$ is described by the environmental model~$\Henv$ that allows nonzero values of the dephasing parameter~$\dfk$, versus (ii) $d$ is described by the vacuum model~$\Hvac$ where no additional parameters are required. While a detectable environmental effect should result in positive evidence for the environmental model, the latter could also be caused by some (hypothetical) non-environmental effect deforming the waveform at the same PN orders. We estimate the magnitude of the environmental effects by computing the marginalized posterior probability distribution of the~$\dfk$ parameter within $\Henv$. 
The posterior distribution is obtained using Bayes' rule for a given prior distribution of the model parameters. 
The hypotheses share the same prior distribution for their common parameters, and we consider a (uninformative) zero-centred uniform prior distribution for~$\dfk$ in~$\Henv$. 
As we discussed in the Supplemental Material, the mismatch between vacuum and environment waveforms shows that the sensitivity of the dephasing parameters for low-mass systems is significantly higher than for a high-mass system---a suitable prior range of $\dfk$ is crucial to ensure the sampler convergence on the global maximum. Throughout this work, we choose to vary a single phase parameter at a time to address specific environmental effects leading to phase deformation.

{\bf \em Density constraints from LIGO-Virgo data.} 
%
To measure the evidence for environmental effects in the data~\cite{GWTC-1_data_release, LIGOScientific:2019lzm, Vallisneri:2014vxa}, from the LIGO~\cite{LIGOScientific:2014pky, LIGOScientific:2020ibl} and Virgo~\cite{VIRGO:2014yos} detectors we analyze individually each of the eleven events of GWTC-1 and two low-mass events from GWTC-2. Table~\ref{tab:bayesF} shows the values of~$\lnB$ for environmental effects with~$k=\{-9,-11\}$ relative to the vacuum model.
We find negative values of the log Bayes factor for all events, which concludes that there is no support for environmental effects in the data.

On the other hand, we can not derive any evidence for the vacuum model either. Environmental corrections are more effective in the early inspiral phase of the coalescence, where LIGO-like detectors might be less sensitive. Even if a binary evolves in a medium, the detection of the environmental effect might be challenging because of noise domination. Nevertheless, we can use the results of this analysis to estimate the upper limits on the environment density using the posterior samples of~$\dfk$, together with the ones of the binary masses.

\begin{figure}[t]
    \centering
    \includegraphics[width=0.95\columnwidth]{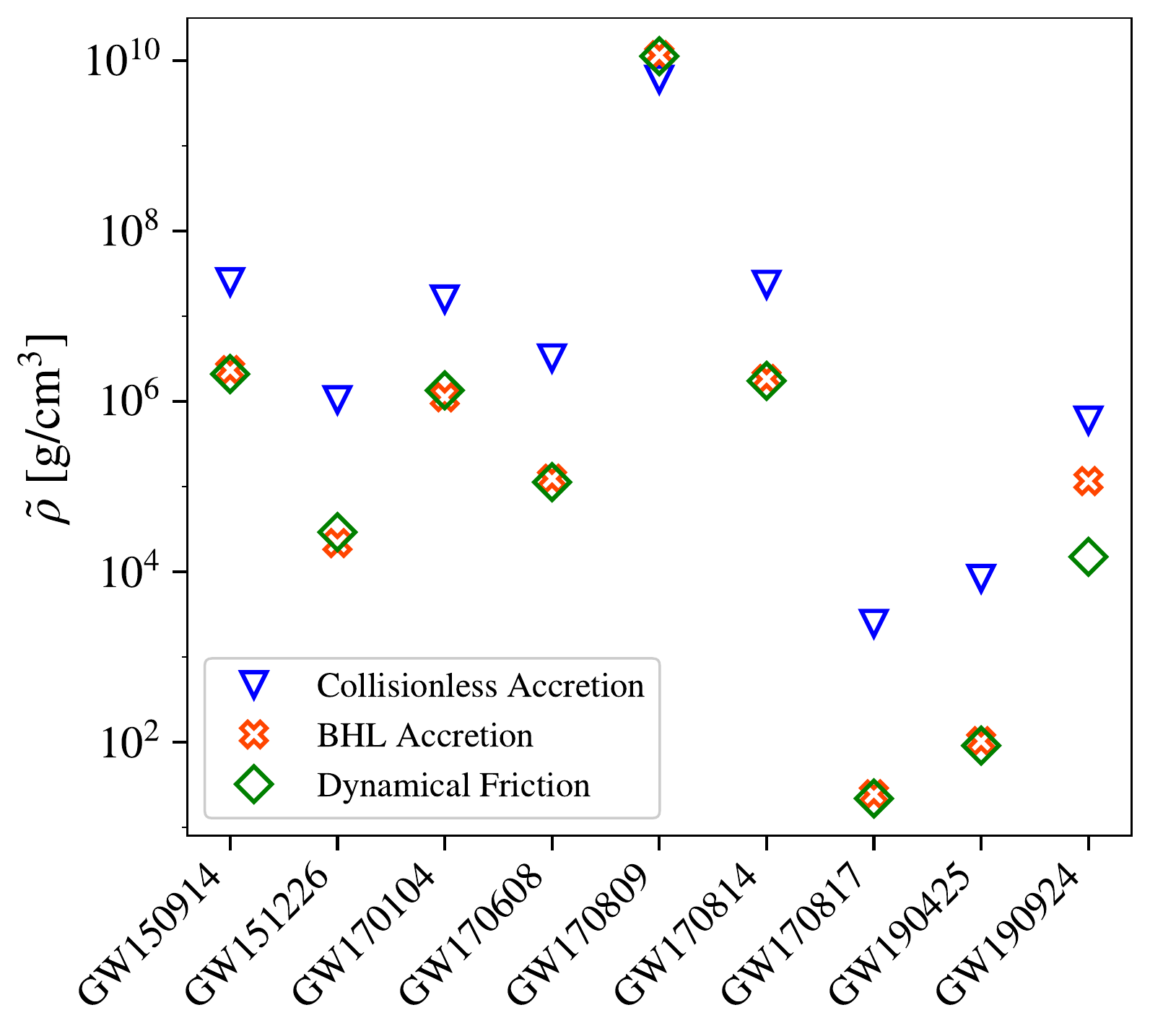}
    \caption{90\% upper bounds on the environmental density obtained considering the effect of CA (blue triangle), BHLA (red cross), and DF (green diamond).}
    \label{fig:density}
\end{figure}

The environmental effects are responsible for negative values of~$\dfk$, but the sampler is not limited to those as we also intend to conduct a model-independent (agnostic) vacuum GR test (see \textit{Supplemental Material}). To put upper bounds on the density, we select only the samples with negative~$\dfk$ values, and calculate the density for the individual environmental effects using~$\dfkhat=\dfk$ and the expressions in Tab.~\ref{tab:coeff}~\footnote{This procedure can have a non-trivial prior effect, we re-weight the posterior to enforce a uniform density prior. We do not find any noticeable effect in this process.}. Figure~\ref{fig:density} shows the 90\% upper bound on the density posterior obtained considering specific environmental effects. Our results for all the events except GW170809 show the density constraint~$\tilde{\rho} \lesssim (21\textup{--}2\times 10^6)\:\densityUnit$, which decisively rules out the binary formation scenario of dynamical fragmentation~\cite{Loeb:2016fzn,Fedrow:2017dpk}. The remaining events' low inspiral SNR with fewer inspiral cycles in the band led to poor density constraints---inconclusive for that scenario. 
We find the notable bound~$\tilde{\rho} \lesssim 21\:\densityUnit$ from GW170817, which is the foremost constraint from our analyses (roughly the density of gold at room temperature on Earth).

\begin{figure}[t]
    \centering
    \includegraphics[width=0.95\columnwidth]{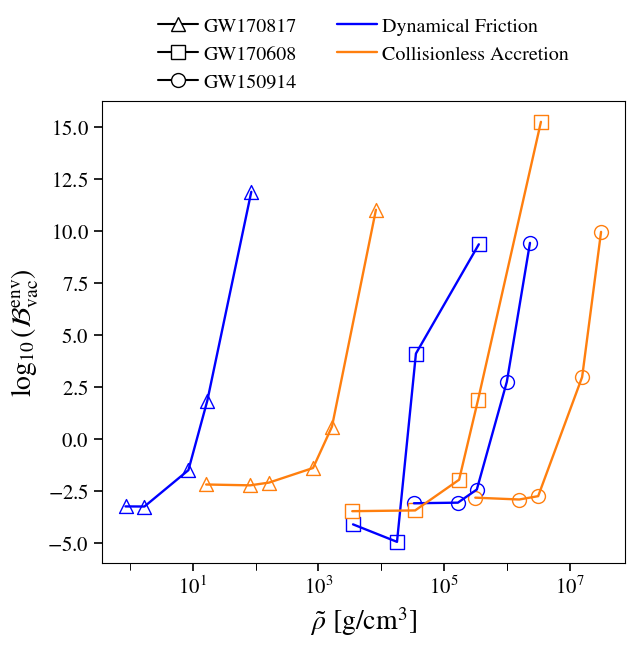}
    \caption{Logarithmic Bayes factor as function of the medium density. This shows the detectability of different environmental effects using the $\aligo$ sensitivity curve, for GW150914, GW170608, and GW170817-like injections in zero noise realization. We omit the BHLA curve since it follows very closely the DF one.
    }
    \label{fig:injection}
\end{figure}
{\bf \em Forecasts for future observations.} 
To understand the detectability of environmental effects with future detectors, we carry out a set of injection analyses mimicking events like GW150914, GW170608, and GW170817 using the design sensitivity of advanced LIGO ($\aligo$)~\cite{aLIGO_ZDHP}, and perform a mismatch-based prediction for Einstein Telescope (ET)~\cite{Hild:2010id} and B-DECIGO~\cite{Isoyama:2018rjb}, with the corresponding lower cutoff frequencies set to~$15$\,\Hz, $5$\,\Hz, and $0.1$\,\Hz, respectively. We select the maximum likelihood sample from the standard vacuum model analysis on real data from the $\bilby$ GWTC-1 rerun~\cite{Romero-Shaw:2020owr, Bilby_GWTC-1_rerun}. Then, we add a set of $\dfk$ values to manipulate the individual environmental effects with a monotonically increasing density parameter. 

\begin{figure}[t]
    \centering
        \includegraphics[width=\columnwidth]{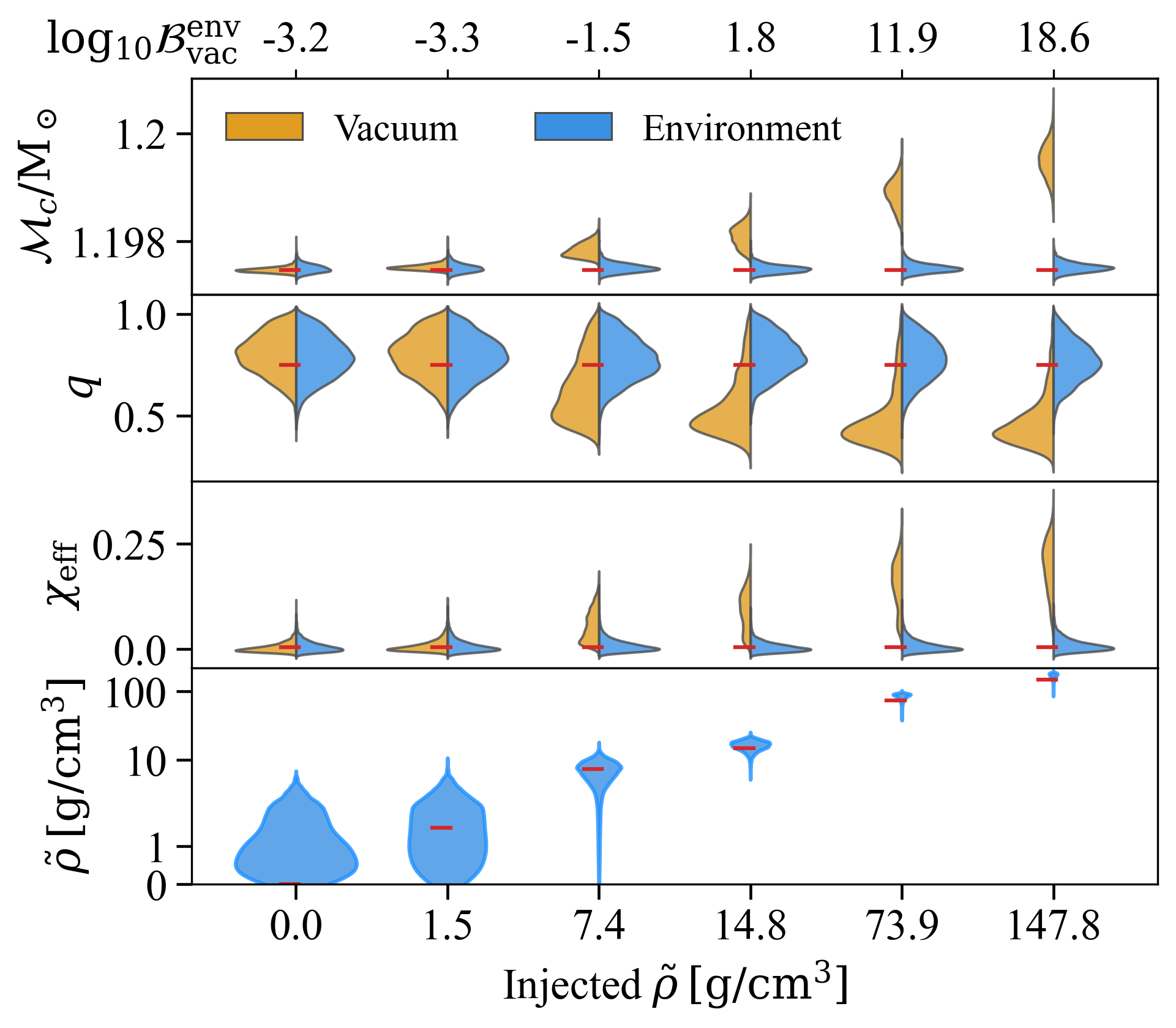}
    \caption{Posterior distribution of the chirp mass ($\mathcal{M}_c$), mass ratio ($q$), and effective spin ($\chi_{\rm eff}$) using the models $\Hvac$ and $\Henv$ with nonzero $\delta\Phi_{-11}$. We injected GW170817-like waveforms deformed by the effect of DF for a set of environment densities (shown in bottom $x$-axis), using the $\aligo$ sensitivity curve. The top $x$-axis shows the logarithmic Bayes factor values for $\Hvac$ versus $\Henv$. Small red marks indicate the injection parameter values.}
    \label{fig:bias}
\end{figure}

For $\aligo$, we conduct Bayesian analysis with zero-noise realization considering a fixed injection SNR ($\rho_{\rm inj}$) of 25 and compare the environment versus vacuum models.
Figure~\ref{fig:injection} shows the values of $\lnB$ for a range of medium densities $0.1$--$10^8\:\densityUnit$. The effect of DF in GW170817, GW170608, GW150914-like events is detectable when $\tilde{\rho} \gtrsim (10,\, 4\times10^4,\, 10^6)\:\densityUnit$, respectively. The effect of CA is detectable only for media~$\sim10$--$100$ times denser. 
When the environment density is set to be half of the (threshold) detectable value, it can still be measurable even though it is not detectable. This is because in comparing environment versus vacuum, the environment hypothesis is hit by Occam's razor factor due to the extra dephasing parameter in $\Henv$. That also explains why almost all $\lnB$ values in Tab.~\ref{tab:bayesF} are negative. 
Figure~\ref{fig:bias} shows that, for the case of the non-vacuum GW170817-like injection, the vacuum model analysis leads to an overestimation of the chirp mass and effective spin, while it underestimates the mass ratio.
Previous studies \cite{2023arXiv230803250L, Chen:2020iky} also discussed the detectability of environmental effects and the biases on the inferred parameters of the binary through the study of the (multi-modal) ringdown phase.
The use of environmental waveform models in future parameter estimation analyses could potentially mitigate this bias.

\begin{figure}[t]
    \centering
    \includegraphics[width=\columnwidth]{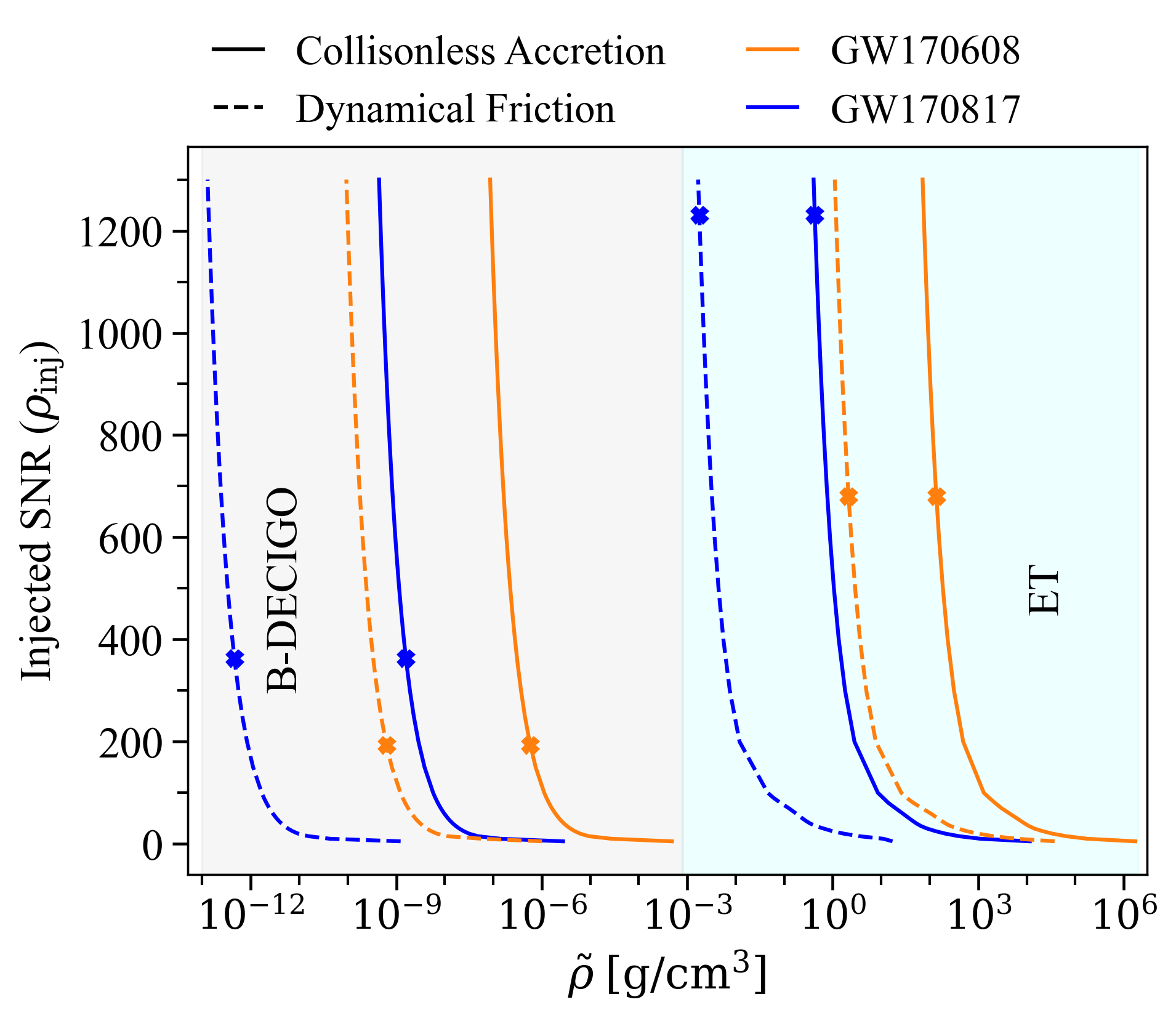}
    \caption{Curves of required SNR for a given density value to achieve $\lnB=3$ for a specific environmental effect, in the configuration of the third-generation detector ET (cyan shade) and the Japanese space-detector B-DECIGO (gray shade). The dots represent the expected SNR if we replace the LIGO-Hanford detector with those future detectors. We omit the BHLA curve since it follows very closely the DF one.
    }
    \label{fig:prediction}
\end{figure}

To study the observability of environments with future detectors like ET and B-DECIGO we use an alternative approach. The Bayes factor between two competitive models can be approximated as~$\ln{\mathcal{B}_{\rm vac}^{\rm env}} \approx \rho^2_{\rm inj} (1 - \ff^2)/2$ when $\Henv$ is true~\cite{Cornish:2011ys}, where $\ff$ refers to the fitting factor of the injected waveform ($h_{\rm inj} \in \Henv$) with the vacuum waveform, such that the match between $\hvac$ and $h_{\rm inj}$ is maximized over all the parameters in $\Hvac$. We assume $\ff\approx 1-\mismatch$ ignoring the correlation between $\dfk$ and the physical parameters in $\Hvac$. Finally, $\mismatch$ is a function only of the dephasing parameter $\dfk$, which in turn is determined by the environment density and the binary parameters through the expressions in Tab.~\ref{tab:coeff}. Figure~\ref{fig:prediction} shows the curves of injected SNR versus environment density necessary to achieve~$\lnB=3$ for events like GW170817 and GW170608. 
Our results suggest that ET will be sensitive to the effect of DF in a GW170817-like event for an environment with~$\tilde{\rho}\gtrsim 10^{-3}\: \densityUnit$ and to CA effects in a medium $\sim10^3$ times denser. As could be already anticipated, due to its better low frequency coverage, B-DECIGO will be sensitive to much lower densities; DF effects in a GW170817-like event will be detectable for an environment with~$\tilde{\rho}\gtrsim 10^{-12}\: \densityUnit$ and CA effects for a density~$\sim 10^4$ times larger.

A forecast based on the Fisher information matrix was previously performed in Ref.~\cite{Cardoso:2019rou}. Overall, our results are more pessimistic (by a few orders of magnitude) than the Fisher analysis: we find a threshold density for detectability larger by roughly 2 orders of magnitude for $\aligo$ and ET; the difference is larger for DECIGO, because we consider the 2030s B-DECIGO.

\vspace{2mm}
{\bf \em Discussion and Conclusions.} 
%
We have designed a model-agnostic Bayesian analysis for detecting the existence of environments surrounding compact binaries. For that, we modelled the waveform considering environmental corrections at $-4.5$PN and $-5.5$PN associated to the leading effect of accretion and DF at linear order in~$\epsilon\coloneqq\tilde{\rho} M_z^2\ll1$. We checked that higher order terms in~$\epsilon$, which affect the phasing at more negative PN orders, are indeed sub-leading and do not alter our results for the systems we studied (see \textit{Supplemental Material}).
Our analyses with LIGO-Virgo events found \emph{no} evidence for the presence of environments. In addition, we used our Bayesian analysis to derive upper bounds on the densities of LIGO-Virgo binary surroundings.
Our results for most of the events decisively rule out the scenario of dynamical fragmentation of massive stars as a possible formation channel for these events~\cite{Loeb:2016fzn}. Our results for GW150914 are also in excellent agreement with the findings of Ref.~\cite{Fedrow:2017dpk}, which used numerical relativity waveforms in gaseous environments.

Even if a compact binary is found evolving in a (low-density) medium, the detection of environmental effects with LIGO-like detectors is challenging, since such detectors are not very sensitive to the early inspiral stage, and the extra dephasing parameter incurs the Occam's razor penalty, resulting in less evidence for the environment model. We found that a medium density can actually be measurable, even when its value is half of the threshold for detectability ($\lnB \gtrsim 0$). We also show that environmental effects can substantially bias the recovered parameters of the vacuum model, even when they are not detectable. Finally, our zero-noise injection analyses indicate that LIGO-like detectors will be capable to see environments (at least) as dilute as~$\tilde{\rho}\sim 10 \: \densityUnit$---roughly the density of lead at room temperature on Earth. While they may in the future definitively exclude the dynamical fragmentation scenario, environments like accretion disk and dark matter overdensities seem to be out of reach.

We have also analysed the prospects for future (2030s) detectors like ET and B-DECIGO. We demonstrated that ET will be sensitive to the effects of DF and BHLA for media as dilute as~$\tilde{\rho}\sim 10^{-3}\: \densityUnit$, implying that it may be capable of detecting environmental effects on a compact binary merging within dense (thin) accretion disks or superradiant clouds hosted by a SMBH. Our results are even more promising for B-DECIGO, indicating that it will be sensitive to DF and BHLA effects from environments as dilute as~$\tilde{\rho}\sim 10^{-12}\: \densityUnit$, which covers, e.g.,  most accretion disk densities, superradiant clouds of ultra-light bosons, or cold dark matter spikes.

In this work we performed a model-agnostic analysis since our goal was to assess the overall capability of LIGO-like and near-future 2030s detectors in probing environments and to derive the first (order of magnitude) constraints on environments from current observations. This also justifies our (unphysical) choice for considering particular environmental effects separately. 
In future work, we plan to focus on specific environments (like accretion disks and superradiant clouds), considering simultaneously several environmental effects entering at different PNs, and to study the distinguishability between different environments with compact binaries (see, e.g., the study of Ref.~\cite{Cole:2022yzw} for EMRIs). We also plan to extend our modelling to asymmetric binaries. Our current studies focused on equal mass binaries and analyzed them using a quadrupolar waveform model. To analyze the events in GWTC-2/3 containing many asymmetric binaries, we need to include the contribution of higher multipoles, which become more relevant with increasing mass ratio. In a follow-up work, we will focus on building an environmental waveform model including higher harmonics. 

\vspace{0.3 cm}
\begin{acknowledgments}
We gratefully acknowledge comments and feedback from Nathan Johnson-McDaniel and Stefano Rinaldi. We are grateful to Marc Andrés-Carcasona for useful discussions throughout the preparation of the paper. We also thank the anonymous referee for the useful comments and suggestions. This research has made use of data or software obtained from the Gravitational Wave Open Science Center (gwosc.org), a service of the LIGO Scientific Collaboration, the Virgo Collaboration, and KAGRA. This material is based upon work supported by NSF's LIGO Laboratory which is a major facility fully funded by the National Science Foundation, as well as the Science and Technology Facilities Council (STFC) of the United Kingdom, the Max-Planck-Society (MPS), and the State of Niedersachsen/Germany for support of the construction of Advanced LIGO and construction and operation of the GEO600 detector. Additional support for Advanced LIGO was provided by the Australian Research Council. Virgo is funded, through the European Gravitational Observatory (EGO), by the French Centre National de Recherche Scientifique (CNRS), the Italian Istituto Nazionale di Fisica Nucleare (INFN) and the Dutch Nikhef, with contributions by institutions from Belgium, Germany, Greece, Hungary, Ireland, Japan, Monaco, Poland, Portugal, Spain. KAGRA is supported by Ministry of Education, Culture, Sports, Science and Technology (MEXT), Japan Society for the Promotion of Science (JSPS) in Japan; National Research Foundation (NRF) and Ministry of Science and ICT (MSIT) in Korea; Academia Sinica (AS) and National Science and Technology Council (NSTC) in Taiwan.
The authors are grateful for computational resources provided by the LIGO Laboratory and supported by National Science Foundation Grants PHY-0757058 and PHY-0823459. S.R. and M.H. were supported by the research program of the Netherlands Organization for Scientific Research (NWO).
R.V. is supported by grant no. FJC2021-046551-I funded by MCIN/AEI/10.13039/501100011033 and by the European Union NextGenerationEU/PRTR. R.V. also acknowledges support by grant no. CERN/FIS-PAR/0023/2019.
This paper has been given LIGO DCC number LIGO-P2300301. This work is partially supported by the Spanish MCIN/AEI/ 10.13039/501100011033  under
the grants SEV-2016-0588, PGC2018-101858-B-I00, and PID2020-113701GB-I00 some of which include ERDF funds from the European  Union. IFAE  is partially funded by the CERCA program of the Generalitat de Catalunya. To generate the waveforms, we have used the~$\lalsim$ package of the LIGO Algorithms Library (LAL) software suite~\cite{lalsuite}. To perform the Bayesian parameter estimation analyses, we have used the nested sampling algorithm~\cite{Skilling:2006gxv, Veitch:2009hd, Speagle_2020} implemented in the $\lalinf$~\cite{Veitch:2014wba} and $\bilby$~\cite{Ashton:2018jfp} packages to evaluate the integral over the model parameter space and calculate the Bayes factor. We have used \numpy~\cite{Harris:2020xlr} and \scipy~\cite{Virtanen:2019joe} for analyses in the manuscript.

\end{acknowledgments}

\bibliography{bibliography}{}

\newpage

\section{Supplemental Material}

\subsection{Dephasing by DF and accretion}
\label{sec;dephasing}
Consider a quasi-circular BBH evolving in some non-vacuum environment (e.g., an AGN accretion disk) where it may have formed. We focus on a stage when the binary is sufficiently close to merger that the GW emission drives the inspiralling, while the environment is responsible only for a subdominant dephasing in the waveform. This regime is consistent with our assumption of quasi-circular orbits, since, even though environmental effects may lead to increase BBH eccentricity~\cite{Cardoso:2020iji,Ishibashi_2020}, the GW radiation reaction circularizes them~\cite{Peters:1964zz}. In the inspiral of a LVK binary of total mass~$M$, the separation distance evolves over a length-scale~$\sim 10^3 \textrm{\,km}\,(M/60 M_\odot)^{1/3}$, which is smaller than the typical length-scale of all environments we are interested in this work (e.g., the size of AGN accretion disks is~$\sim 10^{11} \textrm{\,km}$~\cite{Guo:2022}). Thus, we assume that the BBH is not able to probe the medium density profile, and effectively evolves on a \emph{static uniform} density environment. By assuming a static density profile, we are neglecting the feedback of the BBH on the environment (see, e.g.,~\cite{Kavanagh:2020cfn}); this is admittedly a strong assumption, since the binary is expected to dynamically heat and deplete its surroundings, and may even create a ``cavity'' devoid of matter (as it happens in circumbinary disks~\cite{Rowan:2022ehz}). Nevertheless, if BBHs are formed close enough to merger and do not describe too many orbits, the (average) density of the local environment may remain considerably unchanged (as seen in Ref.~\cite{Fedrow:2017dpk}).

We now follow the same approach as Ref.~\cite{Barausse:2014tra} to compute the leading PN contribution of accretion and DF to the dephasing of the GW signal with respect to the vacuum waveform~\footnote{These are expected to be the most important environmental effects for LVK compact binaries~\cite{Barausse:2014tra,Fedrow:2017dpk,Cardoso:2020lxx}.}. At leading order, the rate of energy loss into GWs is given by the quadrupole formula~\footnote{As in the main text, we use units with~$G=c=1$.}
\begin{equation}
    \dot{E}_{\rm{GW}}\approx\frac{32}{5} \eta^2 v^{10},
\end{equation}
using also Kepler's third law. For close to equal-mass binaries, the rate of energy loss due to the (tangential) DF is given by Eq.~1 in the main text.
Now, at leading (Newtonian) order, the total orbital energy is
\begin{equation}
    E_{\mathrm{orb}}=-\frac{1}{2}\eta M v^{2},
\end{equation}
so, from conservation of energy~$\dot{E}_{\mathrm{orb}}=-\dot{E}_{\rm{GW}}-\dot{E}_{\rm{DF}}$, we arrive at the relation 
\begin{equation}
    \dot{f}\approx \frac{96 \pi^{\frac{8}{3}}}{5} \eta M^{\frac{5}{3}}f^{\frac{11}{3}}+12\Big(\frac{1-3 \eta}{\eta^2}\Big) \rho \mathcal{I} +\frac{5}{2}\frac{f \dot{M}}{M},
\end{equation}
where we used that~$\dot{\eta}\approx 0$ for close to equal-mass binaries. For~$\dot{M}$ we consider either BHL or collisionless accretion. In the case of collisional media, LVK binaries are expected to have supersonic orbital velocities larger than their center of mass velocity with respect to the medium, so that their individual BHL radii are typically comparable to the binary separation distance. Led by numerical simulations
(e.g., \cite{Comerford:2019,Antoni:2019pgq}, but note~\footnote{In these simulations the BHL radius was associated with the center of mass velocity, since the binaries considered had orbital velocities smaller than (or comparable to) the center of mass or the sound velocity.}), we assume that each binary component evolves through BHL accretion, which results in the total mass growth rate
\begin{equation}
    \dot{M}_{\rm{BHLA}} \approx \frac{4 \pi \rho M^2\lambda}{v^3}\Big(\frac{1-5\eta(1-\eta)}{\eta^3} \Big), 
\end{equation}
with the fudge parameter~$\lambda\sim O(1)$. In media constituted by particles with larger mean free path, accretion is less efficient and it is better described by collisionless accretion~\cite{Eddington:1988,Begelman:1977}.
\begin{equation}
    \dot{M}_{\rm{CA}}\approx \frac{16 \pi \rho M^2}{v} \Big(\frac{1-3 \eta}{\eta}\Big).
\end{equation}

\begin{figure}[t]
    \centering
    \includegraphics[width=0.45\textwidth]{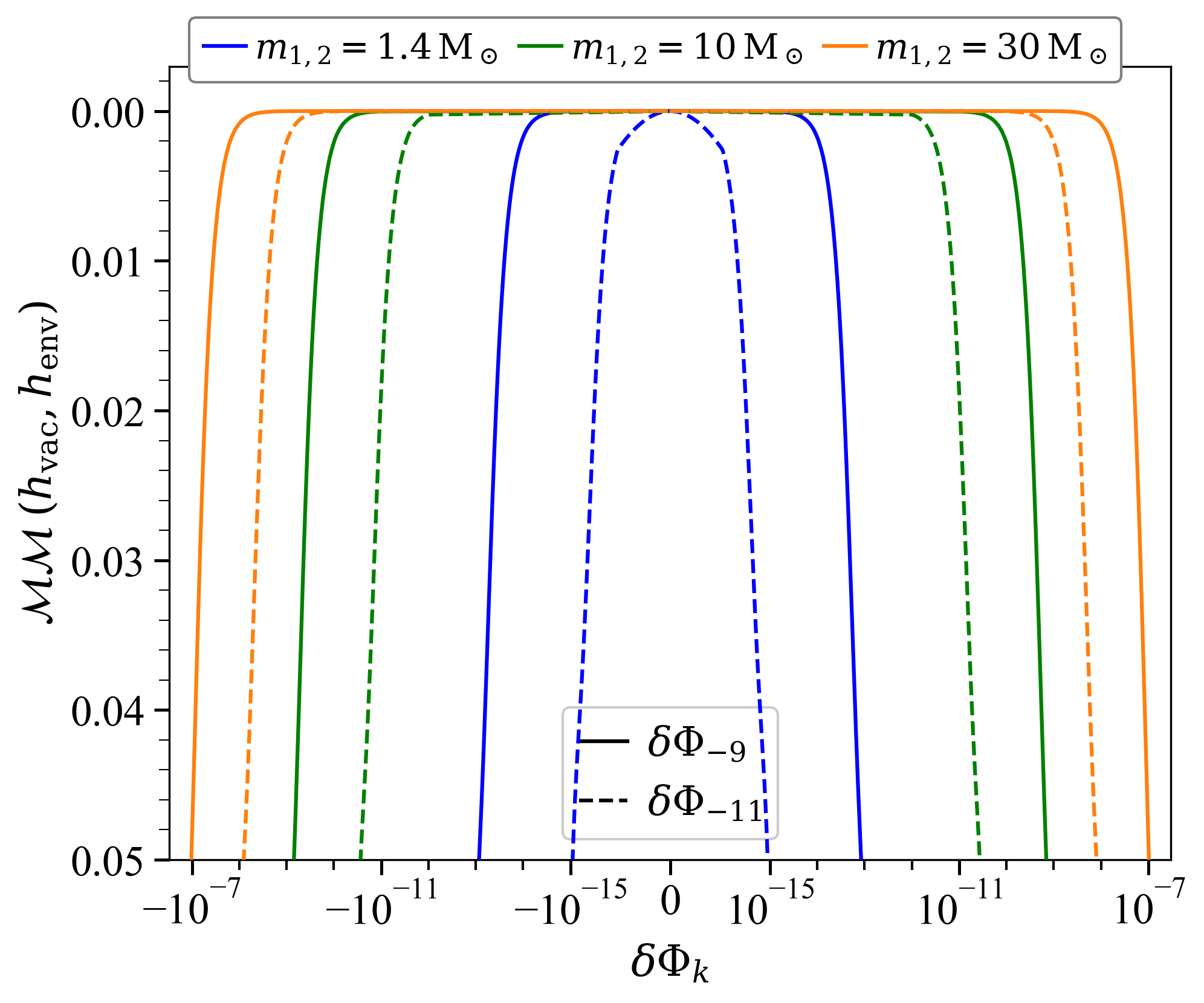}
    \caption{Mismatch ($\mismatch$) between vacuum waveform ($\hvac$) and environment waveform ($\henv$) as function of the dephasing parameter ($\dfk$) for a set a non-spinning equal-mass binary systems. The solid line refers to the effect of CA ($k=-9$), and the dashed line to BHLA or DF ($k=-11$).}
    \label{fig:match}
\end{figure}

Using the stationary phase approximation, we can write the Fourier transform of the plus and cross polarization waveforms as~\cite{Cutler:1994ys}
\begin{equation}
    \tilde{h}_{+,\times} = \mathcal{A}_{+,\times}(f)e^{i \phi_{+,\times}(f)},
\end{equation}
where at leading order (and for the mode~$m=2$),
\begin{gather}
    \mathcal{A}_{+,\times}\approx \frac{Q_{+,\times}}{D}\eta^{\frac{1}{2}}M^{\frac{5}{6}}f^{-\frac{7}{6}},\\
    \phi_+\approx 2\pi f t(f)- \varphi(f)-\frac{\pi}4,\\
    \phi_\times = \phi_+ + \frac{\pi}{2}, 
\end{gather}
with~$Q_{+,\times}(\iota)$ a real function of the binary inclination~$\iota$, and where~$D$ is the binary distance, and~$f$ is the GW frequency. The~$t(f)$ and~$\varphi(f)$ are roughly the instant and the phase when the signal has a frequency~$f$; more precisely, they are defined as
\begin{align}
    t(f)&\coloneqq t_\mathrm{c}-\int^{+\infty}_{f} \mathrm{d}f'\frac{1}{\dot{f}'} , \label{eq:t(f)} \\
    \varphi(f) &\coloneqq \varphi_{\mathrm{c}}-\int^{+\infty}_f \mathrm{d}f'\, \frac{2\pi f'}{\dot{f}'} \label{eq:vphi(f)}.
\end{align}

\begin{figure}[t]
    \centering
    \includegraphics[width=0.45\textwidth]{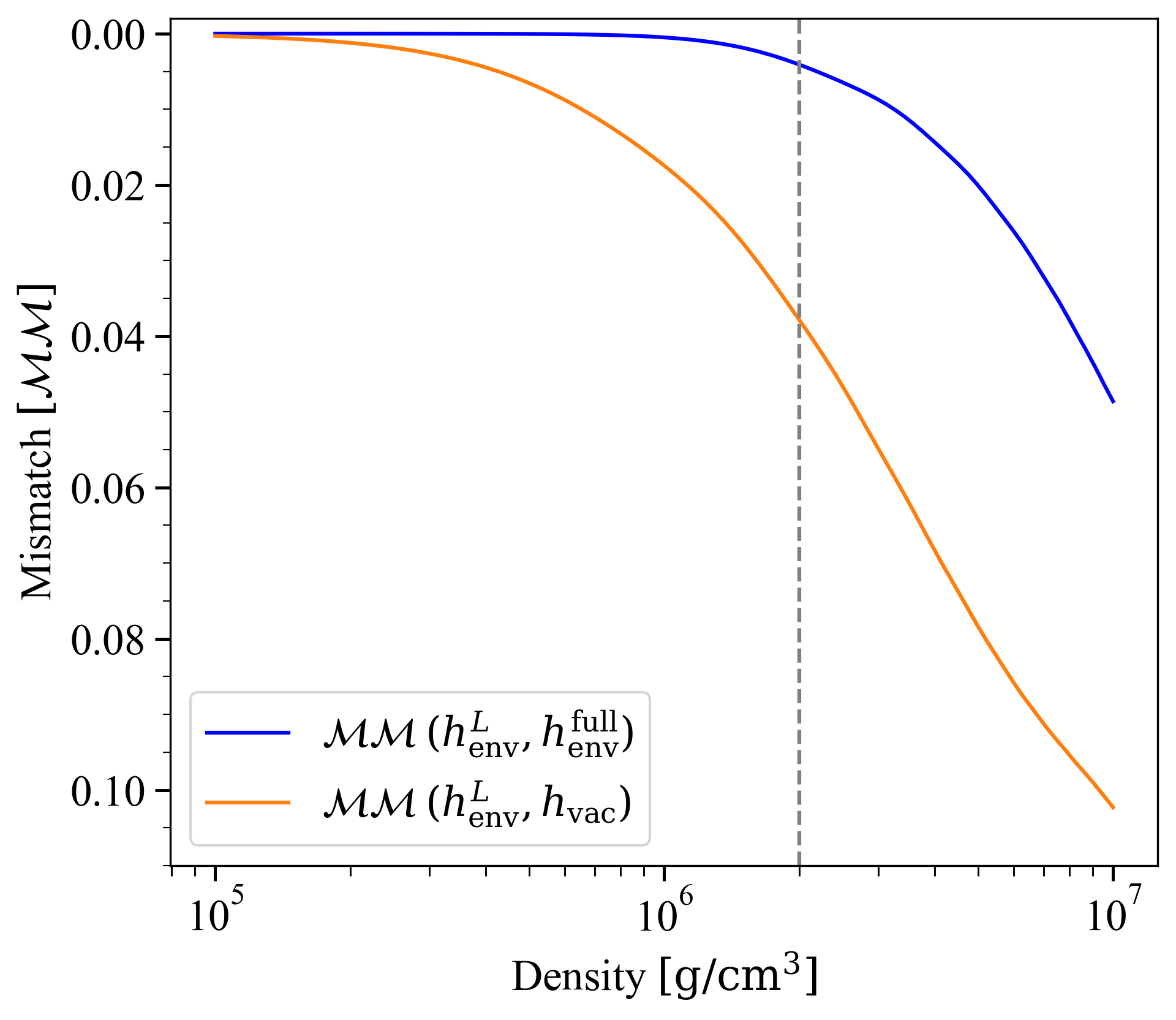}
    \caption{Mismatch as a function of the medium density for a GW150914-like event using the actual GW150914's PSD. The vertical line shows (approximately) the $90\%$ upper bound from Fig.~\ref{fig:density} for DF and BHLA. The $\henv^L$ and $\henv^{\rm full}$ denote, respectively, the environment waveform truncated at linear order in $\rho M^2$ and the full one. We consider simultaneously the environmental effects from DF and BHLA.}
    \label{fig:full_vs_leading}
\end{figure}

In our regime of interest, in which the environmental effects are much smaller than the gravitational radiation-reaction, DF and accretion are responsible only for a slow dephasing of the waveform with respect to vacuum, which, at linear order in $\rho M^2 \ll 1$, can be expressed as
\begin{align}
    \phi_+&\approx 2\pi f t_\mathrm{c}-\varphi_c-\frac{\pi}{4} \nonumber\\ &\qquad+\frac{3}{128 \eta \,v^5}\left(1+\delta_\mathrm{vac}^\mathrm{PN}+\delta_\mathrm{DF}+\delta_\mathrm{accr} \right),
\end{align}
where~$\delta_\mathrm{vac}^\mathrm{PN}$ contains the PN vacuum GR corrections, and with the environmental dephasing parameters
\begin{equation}
    \delta_\mathrm{DF}\approx -\frac{25 \pi(1-3\eta)}{304\, \eta^3}\mathcal{I}\rho M^2 v^{-11},
\end{equation}
and
\begin{equation}
    \delta_\mathrm{accr}\approx \begin{cases}
        -\frac{125 \pi [1-5\eta(1-\eta)]}{1824\, \eta^4}\lambda\rho M^2 v^{-11} \quad \textrm{(BHLA)},\\
        -\frac{125 \pi(1-3\eta)}{357\, \eta^2}\rho M^2 v^{-9}\quad \textrm{(CA)},
    \end{cases}
\end{equation}
Thus, DF and BHLA are both responsible for a~$-5.5\textrm{PN}$ correction, while CA for a~$-4.5\mathrm{PN}$ one to the GW phase. 

We quantify the difference between a vacuum waveform ($\hvac$) and an environment one ($\henv$) by computing the \emph{mismatch}, which is defined as
\begin{equation}
    \mismatch\left(\hvac, \henv\right) = 1-\max_{t_{\rm ref}, \varphi_{\rm ref}}\tfrac{\inp{\hvac}{\henv}}{\sqrt{\inp{\hvac}{\hvac}\inp{\henv}{\henv}}},
\end{equation}
with the maximization taken over an overall phase~$\varphi_{\rm ref}$ and time~$t_{\rm ref}$. The bracket denotes the inner-product weighted by the detector noise power spectral density (PSD). 
We considered the $\aligo$ at design sensitivity \texttt{aLIGOZeroDetHighPower}\:\cite{aLIGO_ZDHP},
ET\:\footnote{The updated sensitivity curve for ET is available at \url{https://apps.et-gw.eu/tds/?content=3&r=18213}, we used the file \texttt{ET20kmcolumns.txt}: HF+LF. }, and B-DECIGO\:\cite{Isoyama:2018rjb} (an initial version of DECIGO\:\cite{Yagi:2011wg}), with the corresponding lower cutoff frequencies set to~$15$\,\Hz, $5$\,\Hz, and $0.1$\,\Hz, respectively, while computing the inner product. Figure~\ref{fig:match} shows the mismatch curves for a set of binary systems using the $\aligo$ sensitivity curve, indicating that with a BNS system, the $\aligo$ is sensitive to much smaller dephasing coefficients ($\sim$ 4--6 orders of magnitude smaller) than in the case of BBHs, which is due to the significantly fewer inspiral cycles in-band for the latter system. 
We note that the mismatch curves are symmetric about $\dfk=0$, but the waveforms are not: a negative value of $\dfk$ results in a shorter waveform than in the vacuum model, while the opposite happens for a positive value.

The higher order terms in~$\rho M^2$ contributing to the \emph{full} environment waveform are responsible for more negative PN order deformations than the linear terms that we considered in our analysis, but we have checked that, for the binary systems studied in this work, they will not change significantly our results. 
More precisely, we verified that the mismatch between the environment waveform truncated at linear order in~$\rho M^2$ and the full environment waveform, $\mismatch(\henv^L,\henv^{\rm full})$, is always much smaller than the mismatch between the environment waveform truncated at linear order and the vacuum waveform, $\mismatch(\henv^L,\hvac)$, for densities smaller than (or comparable to) the $90\%$ upper bounds derived in Figure~\ref{fig:density}. 
Figure~\ref{fig:full_vs_leading} shows one such check for a GW150914-like event, considering the simultaneous effect of DF and BHLA. The full waveform deformation (including all powers in~$\rho M^2$) was expressed analytically in terms of hypergeometric functions from the evaluation of integrals~\eqref{eq:t(f)} and~\eqref{eq:vphi(f)}.

\begin{figure*}[t]
    \centering
    \includegraphics[width=\textwidth]{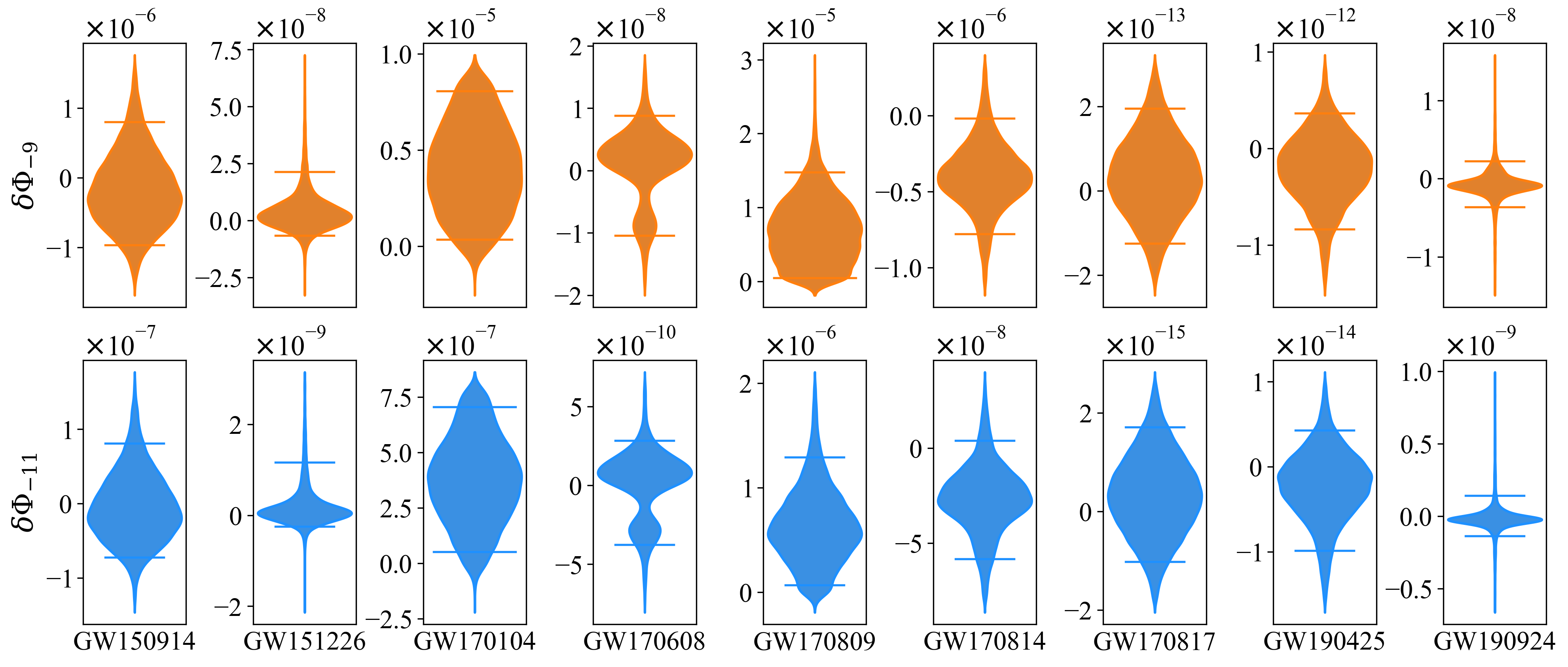}
    \caption{Marginalized posterior distributions for the dephasing parameters $\delta\Phi_{-9}$ (in orange) and $\delta\Phi_{-11}$ (in blue) from the GW observations of compact binaries from GWTC-1 and two low-mass events from GWTC-2, GW190425, and GW190924\_021846. The horizontal lines show the~90\% credible intervals.}
    \label{fig:posteriors}
\end{figure*}

\subsection{Posterior distributions}
\label{sec:posteriors}
We have analyzed each event in GWTC-1 and two low-mass events from GWTC-2 using the on-source data, PSD, and calibration uncertainty provided by the Gravitational Wave Open Science Center (GWOSC)~\cite{GWTC-1_data_release, Vallisneri:2014vxa, LIGOScientific:2019lzm, LIGOScientific:2020ibl}. We have performed the parameter estimation analyses using the $\imrpv$ and $\imrpnrtv$ waveform models for BBH and BNS systems, respectively. We conducted analyses for all LVK events with a lower cutoff frequency of 20Hz, except in the instance of GW170817, where we employed a cutoff frequency of 23Hz. This specific configuration was selected based on information available in the LVK public data. We estimate the magnitude of the environmental effects by computing the marginalized posterior probability distribution of the~$\dfk$ parameter within $\Henv$, integrating the posterior distribution~$p( \vec{\theta}\,|\, d, \Henv)$ over the nuisance parameters,
\begin{equation}
    p\left(\dfk \,|\, d \right) = \int \left( \prod_{\vec{\theta} \,\setminus \{\dfk\}} d\theta_i \right) \: p( \vec{\theta} \,|\, d, \Henv ),
\end{equation}
where~$\vec{\theta}\coloneqq \{\theta_i\}$ is the set of model parameters for~$\Henv$. Our results for the marginalized posterior distributions for the dephasing parameters are shown in Fig.~\ref{fig:posteriors}.

Out of the eleven events analyzed, informative posterior distributions were successfully obtained for nine events, for both $-4.5$PN and $-5.5$PN deformations of the (inspiral) waveform. However, for GW170729 and GW170823, the obtained posterior distributions were uninformative. An uninformative posterior distribution suggests that the data did not impose stringent constraints on the parameters. As a result, the posterior distributions closely resembled the prior distributions. This indicates that the GW data available for these events did not provide sufficient information to make precise inferences about the dephasing parameters. Notably, these events involve binaries with component BHs that stood out as the most massive in the GWTC-1 catalog. Specifically, GW170729 has progenitor masses of $50.2M_{\odot}$ and $24 M_{\odot}$, resulting in the formation of the most massive remnant BH in GWTC-1. Similarly, for GW170823, the component masses were $39.5M_{\odot}$ and $29.0 M_{\odot}$.

In contrast, the BNS system---with the smallest total mass in the catalog---yielded the narrowest distributions for the dephasing parameters: an impressive~$10^{-13}$ and~$10^{-15}$ for $k=-9$ and $k=-11$, respectively. The next highest constraints pertains to the lowest-mass BBH systems: GW151226 and GW170608. These events produced posterior distributions for the dephasing parameters of exceptional precision, with widths of $10^{-8}$ and $10^{-9}$ for $k=-9$ and $k=-11$, respectively. Interestingly, GW170608 shows a faint bimodal tendency. There may be different reasons for this: one factor could stem from the presence of noise below $30$Hz in the Hanford data, which led us to choose a lower cutoff frequency of $30$Hz for H1 and $20$Hz for L1 (the same choice that was made by the LVK).

In the analyses of GW151012 and GW170818 with the environmental model, the sampler failed to converge, in the sense that the individual runs resulted in different posterior distributions. To ascertain the convergence of the posterior, we calculate the Jensen-Shannon (JS) divergence between samples from two distinct runs. This divergence measures how similar the two probability distributions are. A higher value indicates a larger difference between the distributions, suggesting potential issues with convergence. As evaluating the JS divergence is simpler in one dimension, we select it for key parameters: chirp mass ($\mathcal{M}_c$), mass ratio ($q$), effective spin ($\chi_{\rm eff}$), and effective precession spin ($\chi_{\rm p}$). These parameters are crucial for assessing the inconsistency in posterior samples, as they represent significant physical aspects under investigation. We first estimate the probability density functions using the Gaussian KDE from \scipy. To compute the JS divergence between two probability arrays,  we then employ the JS distance metric implemented in \scipy, which is the square root of the JS divergence. We set the JS divergence base to 2, allowing the divergence to range from 0 to 1. To understand the significance of a JS divergence value, we have compared two sets of Gaussian random numbers, each with a unit standard deviation and varying mean values. When the difference in means is larger than 0.2, the JS divergence value exceeds 0.01, and a slight difference becomes visible.
\begin{figure}
    \centering
    \includegraphics[width=0.49\textwidth]{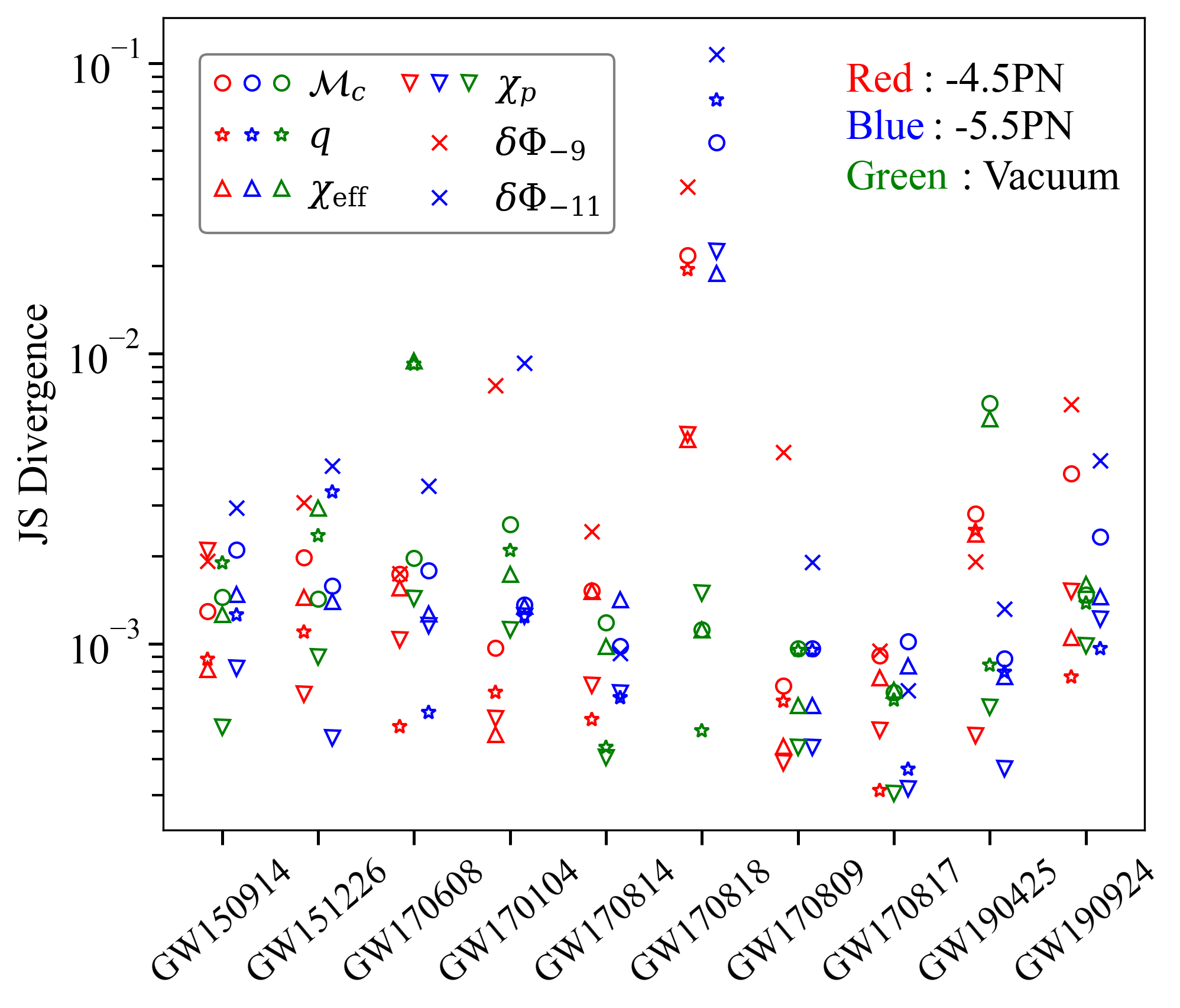}
    \caption{JS divergence between samples from two distinct runs is displayed for three different models: vacuum (green), environmental correction at -4.5PN (red), and environmental correction at -5.5PN (blue).}
    \label{fig:js_divergence}
\end{figure}

Figure~\ref{fig:js_divergence} shows the JS divergence values for the $\mathcal{M}_c$, $q$, $\chi_{\rm eff}$, and  $\chi_{\rm eff}$ posteriors obtained from two distinct runs. It indicates the posteriors from two distinct runs are consistent except for the GW151012 and GW170818 events with the environmental models. To improve the analysis, we reanalyzed these events many times using a better sampler configuration, such as a larger number of live points and autocorrelation times. Unfortunately, the sampler never converged. Like the other high mass events, GW170729 and GW170823, both of these events have a low inspiral SNR and were excluded from parameterized inspiral tests conducted by LVK~\cite{LIGOScientific:2019fpa}.

Despite the large mass of its BBH components, the first observed event (GW150914) shows a unambiguous posterior distribution for both dephasing parameters. This is attributed to the event's high SNR, which compensates for its brief inspiral stage (in band) due to its large mass.
The extent of information that can be derived from each segment of the signal relies on the SNR specific to the stage of coalescence being analysed. In analyses conducted by the LVK collaboration, parameterized inspiral tests are typically employed solely on signals that exhibit an inspiral SNR exceeding 6~\cite{LIGOScientific:2019fpa,LIGOScientific:2020tif,LIGOScientific:2021sio}. However, in this work, we refrain from implementing this SNR threshold. If we were to apply this criterion, it would result in a subset of events comprising GW150914, GW151226, GW170104, GW170608, GW170814, GW170817, GW190425 and GW190924. This indicates that the remaining events have a lower inspiral SNR, posing challenges in extracting information from their signals. 

We have noticed changes in the posterior distributions of the binary source parameters relative to the vacuum analyses. For low-mass systems, the posterior distributions for the chirp mass and effective spin in the environmental model are slightly broader than in the vacuum model, with the peak location of the distributions remaining unchanged. However, for high-mass systems, the changes in the chirp mass and effective spin posteriors are less trivial. When the vacuum model is favored, a broader posterior in the environmental model is expected as the model contains an extra parameter correlated with the chirp mass. This is in agreement with our results for zero-noise injections: looking at Fig.~\ref{fig:bias} closely, particularly in the 1st and 2nd injections (corresponding to small environment densities), one can see that the posterior distributions in the vacuum model are narrower than in the environmental model.


\end{document}